# Ultralow-power coherent qubit control using AQFP logic at millikelvin temperatures

Hiroto Mukai[1,2,3], Akiyoshi Tomonaga[1,2,4], Rui Wang[1,2], Yu Zhou[1], Taro Yamashita[5], Nobuyuki Yoshikawa[6], Jaw-Shen Tsai[1,2], and Naoki Takeuchi[4,7,*]

[1] Center for Quantum Computing, RIKEN, Wako, Saitama, Japan
[2] Graduate School of Science, Tokyo University of Science, Shinjuku, Tokyo, Japan
[3] Department of Applied Physics, The University of Tokyo, Bunkyo, Tokyo, Japan
[4] Global Research and Development Center for Business by Quantum-AI Technology, National Institute of Advanced Industrial Science and Technology (AIST), Tsukuba, Ibaraki, Japan
[5] Graduate School of Engineering, Tohoku University, Sendai, Miyagi, Japan
[6] Institute of Advanced Sciences, Yokohama National University, Yokohama, Kanagawa, Japan
[7] Graduate School of System Informatics, Kobe University, Kobe, Hyogo, Japan

[*] E-mail: takeuchi@port.kobe-u.ac.jp

Qubit controllers are essential for scaling superconducting quantum processors, but implementing them at the 10 mK stage of a dilution refrigerator remains challenging due to stringent cooling constraints. Here we report an ultralow-power qubit controller using adiabatic quantum-flux-parametron (AQFP) logic, termed an AQFP-multiplexed qubit controller with virtual Z gates (AQFP QC-VZ). The AQFP QC-VZ generates multi-tone microwave pulses for qubit control with an ultralow power dissipation of 111 pW per qubit. By combining microwave and time-division multiplexing, the AQFP QC-VZ enables parallel application of X and virtual Z gates to multiple qubits using only a few control lines from room temperature. We demonstrate coherent single-qubit gates at the 10 mK stage using an AQFP mixer, a core component of the AQFP QC-VZ, without observable degradation in coherence.

Superconducting quantum circuits are a leading platform for realizing fault-tolerant quantum computers (FTQCs), as evidenced by recent demonstrations of quantum error correction on a 105-qubit system[1]; however, a much larger number (~$10^6$) of qubits need to be integrated to build a practical FTQC[2–4]. One of the fundamental difficulties in developing large-scale superconducting quantum circuits is the control scheme for multiple superconducting qubits operating at the 10 mK stage of a dilution refrigerator. The maximum number of superconducting qubits that can be manipulated by a conventional control scheme, in which each qubit is controlled by room-temperature electronics through a dedicated coaxial cable, is estimated to be ~$10^3$ because the number of available cables is limited by refrigerator resources such as cooling power and wiring space[5–7]. One promising approach is to incorporate a qubit controller circuit that can manipulate a large number of qubits in the refrigerator; however, this is a significant challenge due to the poor cooling power at the 10 mK stage (typically, 10–100 μW), underscoring the need for ultralow-power qubit controllers.

To this end, various types of qubit controllers have been developed, mainly using cryogenic complementary metal-oxide-semiconductor (cryo-CMOS)[8–12] and superconductor single-flux-quantum (SFQ)[13–17] circuit technologies. Several research groups[8,9,12] have demonstrated qubit control using cryo-CMOS-based qubit controllers, but the controllers were placed at the 3 K stage, rather than the 10 mK stage, because of their relatively large power dissipation (10–100 mW per qubit); thus, the cable count between the 3 K and 10 mK stages remains a major challenge. Acharya *et al.* have developed a very low-power cryo-CMOS-based qubit controller by adopting time-division multiplexing (TDM), i.e., limiting the circuit functionality to microwave demultiplexing, and successfully implemented the TDM controller with qubits at the 10 mK stage. Unfortunately, the TDM controller cannot manipulate multiple qubits in parallel and therefore lacks practicality[9], since in real systems[1,4] many qubits must be controlled simultaneously. As an alternative, other research groups[13–17] have developed SFQ-based qubit controllers using SFQ logic families, such as rapid SFQ (RSFQ)[18] and energy-efficient RSFQ (ERSFQ)[19] circuits, but the power dissipation of these controllers remains insufficient to build large-scale qubit controllers.

Adiabatic quantum-flux-parametron (AQFP) logic[20,21] is an ultralow-power superconductor logic family and is therefore a promising building block for qubit controllers. AQFP circuits can operate with an extremely small power dissipation of ~10 pW per Josephson junction[22] owing to adiabatic switching[23–25], in which the logic state of each AQFP evolves in a quasi-adiabatic way. Previously, we proposed a qubit controller prototype using AQFP logic, referred to as an AQFP-multiplexed qubit controller (AQFP QC)[26]. The AQFP QC comprises an array of cryogenic mixers based on the nonlinearity of AQFPs (referred to as AQFP mixers), enabling the generation of microwave pulses to control multiple qubits in an energy-efficient manner. Although microwave generation tests were successfully demonstrated at 4 K using an AQFP QC chip, the AQFP QC is not functional enough to form a universal quantum gate set. The main problem is that the AQFP QC lacks the capability to execute Z gates, whereas X gates can be implemented by applying the microwave pulses from the AQFP mixers to qubits.

In this study, we report an advanced AQFP-based qubit controller, designated as an AQFP-multiplexed qubit controller with virtual Z gates (AQFP QC-VZ). Whereas the original AQFP QC[26] implements direct mixing of local oscillator (LO) and baseband (BB) signals to generate microwave pulses, the AQFP QC-VZ adopts heterodyne mixing of LO and intermediate frequency (IF) signals to control the phases of microwave pulses and thereby implement virtual Z gates[27]. The AQFP QC-VZ combines microwave multiplexing (μWM)[28,29] and time-division multiplexing (TDM) to apply both X and virtual Z gates to multiple qubits with a few coaxial cables between the 300 K and 10 mK stages. Furthermore, the power dissipation of the AQFP QC-VZ is estimated to be 111 pW/qubit, which is sufficiently small compared with the cooling power at the 10 mK stage. These features highlight the scalability of the AQFP QC-VZ.

We validate the architectural aspects of the AQFP QC-VZ by numerical simulations and 4 K experiments, demonstrating microwave manipulation with an AQFP QC-VZ chip. Furthermore, we verify AQFP-based qubit control by demonstrating single-qubit gates using an AQFP mixer, a core component of the AQFP QC-VZ, as a proof of principle. We demonstrate X and virtual Z gates by applying microwave pulses generated by the AQFP mixer to a transmon qubit at the 10 mK stage. The experimental results on qubit coherence characterization reveal that the presence of the AQFP mixer does not induce observable decoherence to the transmon qubit within our measurement resolution.

## AQFP-multiplexed qubit controller with virtual Z gates (AQFP QC-VZ)

Figure 1a illustrates a conceptual diagram of the AQFP QC-VZ, where the AQFP QC-VZ and qubit chips are interconnected at the 10 mK stage inside a dilution refrigerator. The AQFP QC-VZ chip produces multi-tone microwave pulses for qubit control by mixing a multi-tone LO signal with an IF signal, each provided through a single coaxial cable from the 300 K stage. As a result, the AQFP QC-VZ chip can control a large number of qubits with only two control lines between the 300 K and 10 mK stages. This offers a key advantage over a conventional control scheme, which requires as many coaxial cables between the 300 K and 10 mK stages as there are qubits. The AQFP QC-VZ chip can execute arbitrary single-qubit gates by applying microwave pulses (i.e., X gates) with appropriate phases (i.e., virtual Z gates) to qubits, where the phases are controlled by dynamically varying the IF signal phase. The implementation of two-qubit gates using AQFP logic will be investigated in future work.

Figure 1b depicts a schematic diagram of an AQFP QC-VZ, controlling three qubits as an example. The AQFP QC-VZ consists of an array of AQFP mixers and superconducting resonators (operating as a microwave demultiplexer) for qubit control via μWM. A multi-tone LO signal ($I_{lo}$), in which three microwave tones ($f_1$ through $f_3$) are multiplexed, is provided via a single coaxial cable and demultiplexed by the resonators, assuming the resonator frequencies match the microwave tones. Each AQFP mixer is coupled to one of the resonators and the common IF signal ($I_{if}$), thereby generating microwave pulses with a unique frequency. For example, AQFP mixer 1 is coupled to the $f_1$-resonator and $I_{if}$, thus generating microwave pulses ($V_{out1}$) with a difference frequency of $f_1 - f_{if}$ ($f_{if}$: frequency of $I_{if}$), where the other frequency components ($f_1$, $f_{if}$, $f_1 + f_{if}$, etc.) are omitted by a bandpass filter (BPF) integrated in the mixer. Similarly, AQFP mixers 2 and 3 generate microwave pulses ($V_{out2}$ and $V_{out3}$) with difference frequencies of $f_2 - f_{if}$ and $f_3 - f_{if}$, respectively. Consequently, the AQFP mixers can apply multi-tone microwave pulses to multiple qubits using only two supply signals ($I_{lo}$ and $I_{if}$), where the pulse frequencies are assumed to match the qubit frequencies.

To demonstrate a pulse sequence for a given quantum algorithm, $V_{out1}$ through $V_{out3}$ are individually switched on/off by applying digital input currents ($I_{in1}$ through $I_{in3}$) to the AQFP mixers from a cryogenic memory (not described in this study but intended to be integrated in the AQFP QC-VZ). As shown in Fig. 1c, an AQFP mixer consists of two AQFPs that generate microwave pulses with a relative phase of 0 or π, depending on the digital input currents ($I_{in}$ and $I_{fix}$). Therefore, the output of the AQFP mixer can be switched on/off by the interference between the microwaves from the two AQFPs. Hereafter, $I_{fix}$ is fixed to logic 1, so the output of the AQFP mixer is turned on and off for a positive $I_{in}$ (i.e., logic 1) and negative $I_{in}$ (i.e., logic 0), respectively. Furthermore, to perform virtual Z gates, the phases of $V_{out1}$ through $V_{out3}$ are controlled by dynamically varying the phase of $I_{if}$ ($\theta_{if}$), since their phases are given by $-\theta_{if}$. As shown later, $\theta_{if}$ is changed using TDM to individually set the phase of each pulse.

Figure 1d shows numerically simulated waveforms for the AQFP QC-VZ in Fig. 1b using JoSIM[30], exhibiting the capability to execute arbitrary single-qubit gates. $I_{lo}$ includes three microwave tones ($f_1$ = 8.259 GHz; $f_2$ = 8.514 GHz; $f_3$ = 8.758 GHz), and $I_{if}$ has a triangular envelope for $f_{if}$ = 3.5 GHz and a repetition period (i.e., the period of control cycles) of

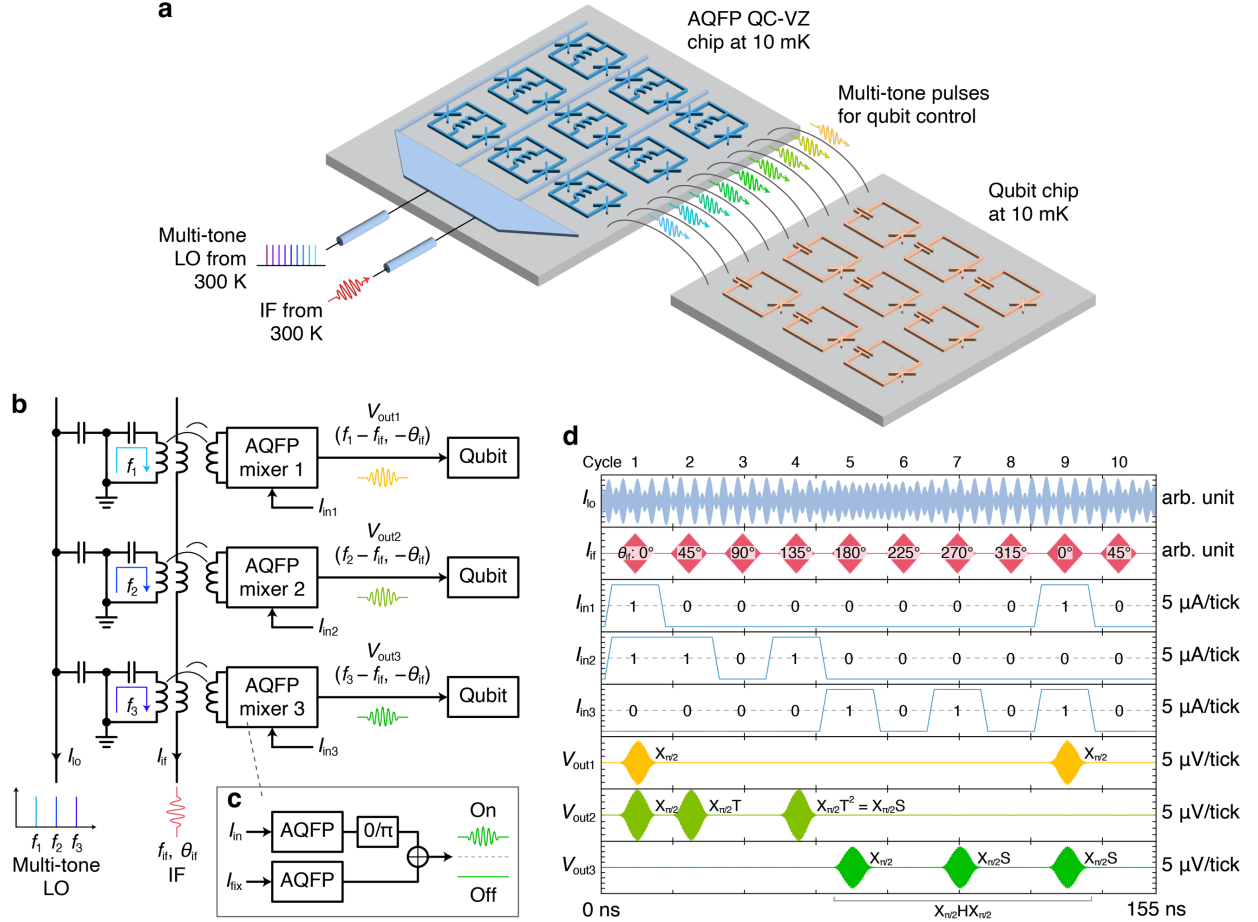

**Fig. 1 | AQFP-multiplexed qubit controller with virtual Z gates (AQFP QC-VZ). a.** Conceptual diagram. An AQFP QC-VZ chip produces multi-tone microwave pulses from multi-tone LO and IF signals and applies them to a qubit chip, thereby controlling multiple qubits with ultralow power dissipation and a few coaxial cables from room temperature. **b.** Detailed circuit schematic, comprising an AQFP mixer array and a resonator array operating as a microwave demultiplexer. Each AQFP mixer generates microwave pulses with a unique frequency by mixing a microwave tone in the LO signal ($I_{lo}$) and the common IF signal ($I_{if}$). The outputs of the AQFP mixers ($V_{out1}$ through $V_{out3}$) are individually switched on/off by digital input currents ($I_{in1}$ through $I_{in3}$) to execute a given pulse sequence. **c.** Block diagram of an AQFP mixer, comprising two AQFPs generating microwave pulses with a relative phase of 0 or π. The output is switched on/off by the interference between the microwaves from the two AQFPs. **d.** Numerical simulation. The IF phase ($\theta_{if}$) changes by 45° at every control cycle to individually set the phase of each microwave pulse. The relative phase between consecutive pulses applies a virtual Z gate. The waveforms at $V_{out1}$ through $V_{out3}$ demonstrate T, S, and H gates, indicating the capability to execute arbitrary single-qubit gates.

15 ns. AQFP mixers generate $V_{out1}$ through $V_{out3}$ of $f_1 - f_{if}$ = 4.759 GHz, $f_2 - f_{if}$ = 5.014 GHz, and $f_3 - f_{if}$ = 5.258 GHz, respectively, in accordance with the pulse sequence given by $I_{in1}$ through $I_{in3}$, where the dashed lines represent zero for each waveform. It is assumed the pulse shapes of $V_{out1}$ through $V_{out3}$ were calibrated such that each microwave pulse would correspond to a π/2 pulse, i.e., $X_{\pi/2}$ gate (the pulse shapes can be individually calibrated by adjusting each microwave tone in $I_{lo}$ and the envelope of $I_{if}$[26]). Importantly, $\theta_{if}$ changes by 45° at every control cycle to individually set the phase of each π/2 pulse via TDM. For $V_{out1}$, two π/2 pulses are generated at the first and ninth cycles. Since $\theta_{if}$ = 0° at both cycles, the relative IF phase ($\Delta\theta$) between the first and second π/2 pulses is 0°. Therefore, each pulse corresponds to $X_{\pi/2}$, i.e., $V_{out1}$ executes a gate sequence of $X_{\pi/2}X_{\pi/2}$. For $V_{out2}$, the first and second pulses are generated at the first ($\theta_{if}$ = 0°) and second ($\theta_{if}$ = 45°) cycles, respectively.

Thus, $\Delta\theta$ = 45°, which virtually applies a $Z_{\pi/4}$ gate (i.e., T gate) between the first and second pulses[27,31]. Similarly, since the third pulse is generated at the fourth cycle ($\theta_{if}$ = 135°), $\Delta\theta$ = 90° between the second and third pulses, virtually applying a $Z_{\pi/2}$ gate (i.e., S gate). As a result, $V_{out2}$ executes a gate sequence of $X_{\pi/2}SX_{\pi/2}TX_{\pi/2}$. For $V_{out3}$, the first through third pulses respectively correspond to $X_{\pi/2}$, $X_{\pi/2}S$, $X_{\pi/2}S$, thus executing a gate sequence of $X_{\pi/2}HX_{\pi/2}$, where H = $SX_{\pi/2}S$ is the Hadamard gate. From the above, the AQFP QC-VZ can execute H, S, and T gates, which form all single-qubit gates in the "Clifford + T" set[32], and can thus apply arbitrary single-qubit gates to each target qubit.

Note that this phase control is not possible in the original AQFP QC. This is because the AQFP QC uses direct mixing of LO and BB signals, and each LO tone is phase-locked by a resonator in the microwave demultiplexer.

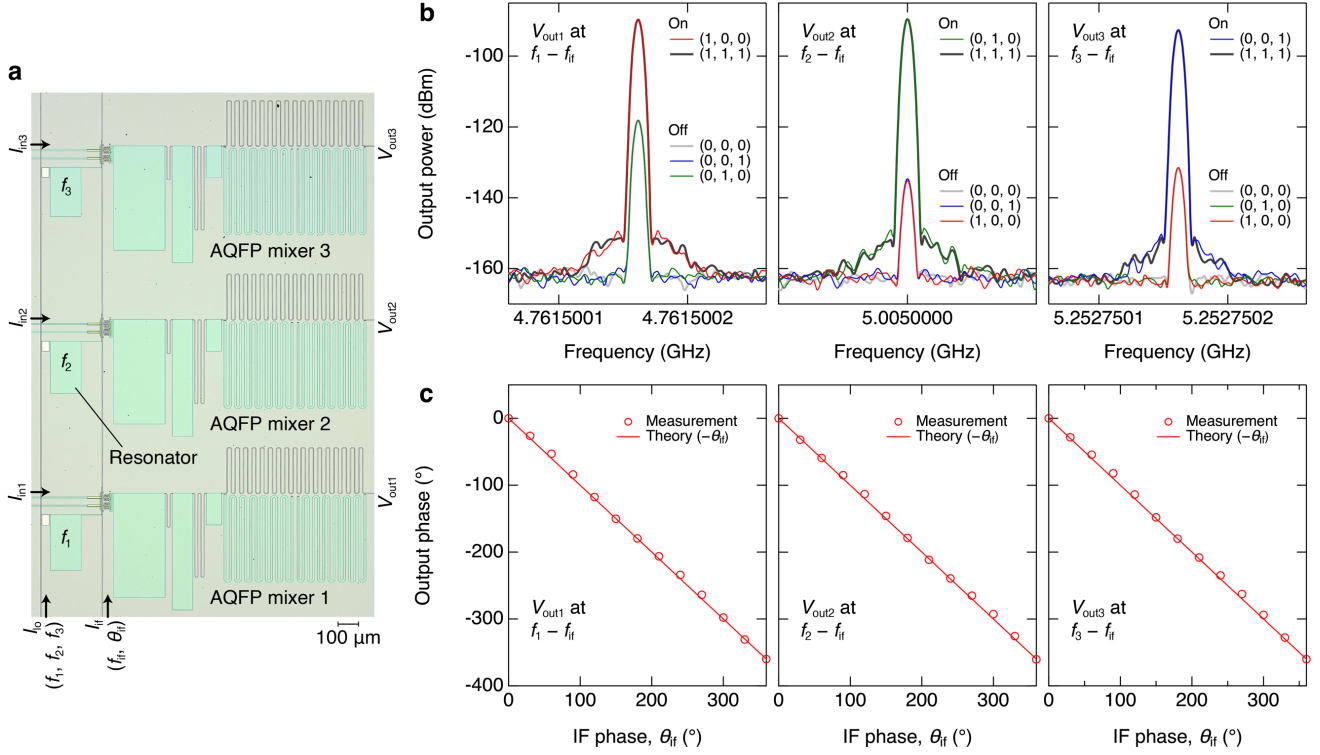

**Fig. 2 | Microwave manipulation using the AQFP QC-VZ. a.** Micrograph of an AQFP QC-VZ chip, comprising three multiplexed AQFP mixers. The mixers generate multi-tone microwave signals ($V_{out1}$ through $V_{out3}$) from a multi-tone LO signal ($I_{lo}$), including $f_1$ through $f_3$ tones, and a common IF signal ($I_{if}$). The outputs of the AQFP mixers are switched on/off by digital input currents ($I_{in1}$ through $I_{in3}$). **b.** Switching operation for $V_{out1}$ through $V_{out3}$ with $I_{in1}$ through $I_{in3}$, the logical values of which are represented by ($a_1$, $a_2$, $a_3$). Each microwave output is generated at a unique difference frequency and is individually switched on/off by an associated input current. For instance, $V_{out1}$ is generated at $f_1 − f_{if}$ ($f_{if}$: IF frequency) and is switched on/off by $I_{in1}$. The on/off ratios of $V_{out1}$ through $V_{out3}$ are 28.5 dB, 45.1 dB, and 39.0 dB, respectively. **c.** Phase control for $V_{out1}$ through $V_{out3}$ with the IF phase ($\theta_{if}$). The phases of all microwave outputs change in the form of $-\theta_{if}$. The above results demonstrate all microwave manipulation capabilities of the AQFP QC-VZ: multiplexing, demultiplexing, mixing, switching, and phase control.

## Circuit performance estimation

The output power and power dissipation of the AQFP QC-VZ were estimated by JSIM_n[33], based on the circuit parameters used for the transient analysis in Fig. 1d, where the critical current density and Stewart-McCumber parameter ($\beta_c$)[34,35] of Josephson junctions were 1 kA/cm² and 25, respectively, assuming the use of the AIST 1 kA/cm² Nb process (1KP)[36]. The maximum output power of the AQFP mixer, when $\Phi_{lo} = \Phi_{if} = 0.5\Phi_0$ ($\Phi_{lo}$: flux amplitude applied to each AQFP by $I_{lo}$; $\Phi_{if}$: that by $I_{if}$; $\Phi_0$: flux quantum), was 4.11 pW (= −83.9 dBm) at 5 GHz. The standby ($\Phi_{lo} = 0.5\Phi_0$; $\Phi_{if} = 0$) and peak ($\Phi_{lo} = \Phi_{if} = 0.5\Phi_0$) power dissipations of the AQFP mixer were 1.92 pW and 220 pW, respectively, at 5 GHz. Defining the average power dissipation of the AQFP mixer as an average of the standby and peak power, the power dissipation of the AQFP QC-VZ is 111 pW per qubit on average.

Here, we introduce control parallelism as a performance metric for evaluating the maximum number of qubits that can be controlled simultaneously within a single control cycle. As shown in Fig. 1d, the AQFP QC-VZ employs TDM-based phase control to apply virtual Z gates to multiple qubits with a single IF signal, significantly reducing the number of cables between the 300 K and 10 mK stages but potentially limiting control parallelism. In the worst case, all AQFP mixers execute gate operations randomly, i.e., each AQFP mixer is randomly excited at eight possible $\theta_{if}$ values (0°, 45°, 90°, …, 315°), thus limiting the average gate parallelism to $N_{qubit}/8$ ($N_{qubit}$: qubit count). In the best case, all AQFP mixers execute the same gate operations and are thus excited at the same $\theta_{if}$ value, skipping the control cycles with unused $\theta_{if}$ values; consequently, the gate parallelism is maximized to $N_{qubit}$. In practice, when implementing a surface code[2], the same gate operations are applied to many qubits in parallel at every control cycle[1], so the practical gate parallelism of the AQFP QC-VZ is expected to be close to $N_{qubit}$.

As discussed in the literature[26], the maximum number of microwave tones that can be multiplexed on a single LO line is determined by the quality factor ($Q$) of the superconductor resonators in a microwave demultiplexer and the LO line bandwidth ($W$). Assuming $Q \sim 10^4$ for Nb resonators[37] and $W$ = 2 GHz, approximately 4000 tones can be multiplexed on the LO line at maximum, i.e., the number of coaxial cables between the 300 K and 10 mK stages can be reduced to $\sim N_{qubit}/4000$. From the above, the AQFP QC-VZ can control multiple qubits with ultralow power dissipation (111 pW per qubit), high control parallelism (~$N_{qubit}$), and a small number of coaxial cables (~$N_{qubit}/4000$) from the 300 K stage.

## Microwave manipulation experiments

To demonstrate microwave manipulation capabilities of the AQFP QC-VZ, we fabricated an AQFP QC-VZ chip using the 1KP[36] and tested it at 4.2 K in liquid helium with a wideband cryoprobe[38]. Figure 2a shows a micrograph of an AQFP QC-VZ chip, comprising three AQFP mixers and a resonator array operating as a microwave demultiplexer. The resonance frequencies of the resonators ($f_1$ through $f_3$) in the demultiplexer were found to be 7.74225 GHz, 7.98575 GHz, and 8.23350 GHz, respectively, from the insertion loss of the LO line measured with a vector network analyzer (Keysight P9374A). During experiments, a multi-tone LO signal ($I_{lo}$), including $f_1$ through $f_3$ tones, and a continuous IF signal ($I_{if}$) with $f_{if}$ = 2.98075 GHz were provided to the chip by an rf arbitrary waveform generator (Keysight M5300A), where the amplitudes of $I_{lo}$ and $I_{if}$ were adjusted such that $\Phi_{lo} = \Phi_{if} = 0.5\Phi_0$. The outputs from the mixers ($V_{out1}$ through $V_{out3}$) were observed by a signal analyzer (Keysight N9010B) and switched on/off by digital input currents ($I_{in1}$ through $I_{in3}$) from a source measure unit (Keysight M9614A).

Figure 2b demonstrates mixing and switching operations of the AQFP mixers, where the output power of each mixer was measured for several combinations of ($a_1$, $a_2$, $a_3$), where $a_1$ through $a_3$ represent the logical values of $I_{in1}$ through $I_{in3}$, respectively. The left panel of Fig. 2b shows that $V_{out1}$ was generated at a difference frequency of $f_1 - f_{if}$ by mixing the $f_1$ tone in $I_{lo}$ with $I_{if}$, and that $V_{out1}$ was turned on for $a_1 = 1$ [i.e., (1, 0, 0) and (1, 1, 1)] and off for $a_1 = 0$ [i.e., (0, 0, 0), (0, 0, 1), and (0, 1, 0)]. Similarly, the center and right panels of Fig. 2b show that $V_{out2}$ and $V_{out3}$ were generated at $f_2 - f_{if}$ and $f_3 - f_{if}$, respectively, by mixing the $f_2$ and $f_3$ tones in $I_{lo}$ with $I_{if}$, and that $V_{out2}$ and $V_{out3}$ were individually switched on/off by $a_2$ and $a_3$, respectively. The output powers at $V_{out1}$ through $V_{out3}$ were −89.7 dBm, −89.6 dBm, and −92.7 dBm, respectively. These power values may have been lower than the simulation results because the current swing in each resonator was suppressed by relatively large dielectric losses in the $SiO_2$ interlayers[39], which may be improved by process modifications in the 1KP (e.g., replacement of the interlayer material[40]). The on/off ratios for $V_{out1}$ through $V_{out3}$ were 28.5 dB, 45.1 dB, and 39.0 dB, respectively, which are comparable to those of previously reported cryogenic microwave switches[41–43].

Figure 2c demonstrates phase control operations of the AQFP mixers, where the output phase of each mixer was measured by converting the microwave output into a dc voltage with an additional mixer[44]. The left-to-right panels in Fig. 2c show that the output phase of each mixer could be controlled in good agreement with the expected linear dependence of −$\theta_{if}$, where each output phase was defined as 0° for $\theta_{if}$ = 0°. The above measurement results demonstrate all microwave manipulation capabilities of the AQFP QC-VZ: multiplexing, demultiplexing, mixing, switching, and phase control.

## Observation of Rabi oscillations

As a first step toward large-scale qubit control using AQFP logic, we demonstrate single-qubit experiments with an AQFP mixer. We implemented a transmon qubit chip and an AQFP mixer chip in separate cryogenic packages and placed them together at the 10 mK stage in a dilution refrigerator. As shown in Fig. 3a, the qubit and AQFP chips were interconnected by a coaxial cable so that the microwave pulses from the AQFP mixer could be transmitted to the transmon qubit. The microwave pulses ($V_{out}$) were generated by mixing an LO signal ($I_{lo}$) and an IF signal ($I_{if}$), with $I_{lo}$ directly applied to the AQFP mixer without a microwave demultiplexer, since this setup was for single-qubit experiments. $I_{lo}$ and $I_{if}$ were provided by a signal generator (Keysight E8257C) and an rf arbitrary waveform generator (Keysight M5300A), respectively. A digital input current ($I_{in}$) was also applied to the AQFP mixer to switch on/off $V_{out}$. See Supplementary Information for more details regarding the experimental setup.

First, we observed Rabi oscillations to verify that AQFP logic can drive qubits coherently. Figure 3b illustrates the signal sequence for this experiment. Following the initialization of $I_{in}$ and $I_{lo}$ with a frequency of $f_{lo}$, a pulsed $I_{if}$ with a duration of $\tau_{if}$ and a frequency of $f_{if}$ was applied to the AQFP mixer. Then, a readout pulse ($V_{read}$) was applied to observe the qubit state. Figure 3c shows the resulting Rabi oscillations as a function of $f_{if}$ when $f_{lo}$ was fixed to 8.0 GHz and the AQFP mixer was turned on by a positive $I_{in}$ of 20 μA. The clear chevron interference patterns indicate the coherent drive of the transmon qubit[13] by the AQFP mixer. Figure 3d shows the results when the AQFP mixer was turned off by applying a negative $I_{in}$ of −20 μA, indicating suppressed Rabi oscillations. The comparison between Figs. 3c and d clearly demonstrates that the observed qubit dynamics were induced by the microwave pulses from the AQFP mixer, rather than residual leakage from the supply signals.

We next performed a quantitative analysis for on/off modulation using the AQFP mixer. $f_{if}$ was fixed to the center frequency (= 3.46798 GHz) of the chevron pattern in Fig. 3c, i.e., the drive frequency was fixed to $f_{lo} - f_{if}$ = 4.53202 GHz, corresponding to the qubit transition frequency ($f_{qubit}$). Then, Rabi frequency ($\Omega_R/2\pi$) was measured as a function of the IF amplitude ($A_{if}$) for both AQFP mixer states (on and off), where each $\Omega_R/2\pi$ value was extracted by fitting the Rabi oscillation to a sinusoidal function. Figure 3e summarizes the extracted $\Omega_R/2\pi$ vs. $A_{if}$, normalized by the maximum amplitude of the waveform generator (~ −2 dBm at 300 K), with the LO power fixed to 10 dBm at 300 K. When the AQFP mixer was on, $\Omega_R/2\pi$ increased monotonically with $A_{if}$, consistent with conventional qubit drive using room-temperature electronics[31]. In contrast, when the AQFP mixer was off, the change of $\Omega_R/2\pi$ with $A_{if}$ was significantly suppressed. To compare the effective drive strength between the on and off states of the AQFP mixer, we define a residual ratio $\Omega_R^{off}/\Omega_R^{on}$, where $\Omega_R^{on}$ and $\Omega_R^{off}$ denote the $\Omega_R$ values for the AQFP mixer on and off, respectively. Figure 3f shows $\Omega_R^{off}/\Omega_R^{on}$ as a function of $A_{if}$. At a high-power region ($A_{if} \gtrsim 0.4$), the residual ratio decreases with increasing $A_{if}$ and reaches approximately 0.05 for $A_{if}$ = 1.0, which may be further improved by amplifying the IF power.

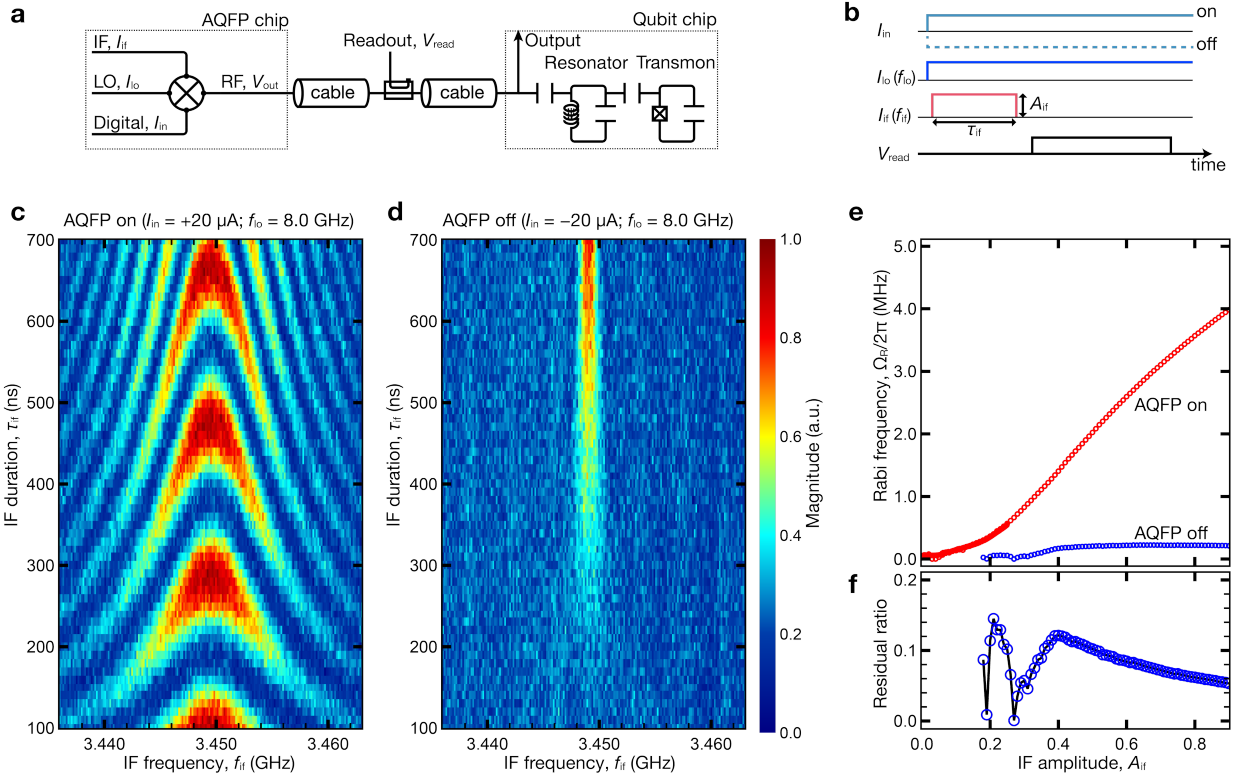

**Fig. 3 | Rabi oscillation experiments using an AQFP mixer. a.** Schematic of the experimental setup. An AQFP mixer generates microwave pulses ($V_{out}$) by mixing an IF signal ($I_{if}$) and an LO signal ($I_{lo}$), which are transmitted to a transmon qubit via a coaxial cable. $V_{out}$ is switched on/off by a digital input current ($I_{in}$). **b.** Timing diagram for $I_{in}$, $I_{lo}$, $I_{if}$, and a readout pulse ($V_{read}$). $I_{if}$ has parameters of a frequency ($f_{if}$), an amplitude ($A_{if}$), and duration ($\tau_{if}$). **c. and d.** Rabi oscillations for both AQFP mixer states (on and off) as functions of $f_{if}$, with the LO frequency ($f_{lo}$) fixed to 8.0 GHz. Clear chevron interference appears when the AQFP mixer is on, whereas the interference is suppressed when the mixer is off. This indicates that the Rabi oscillations were induced by the microwave pulses from the AQFP mixer, rather than the residual leakage from supply signals. From the center frequency of the chevron pattern ($f_{if}$ = 3.46798 GHz), the qubit frequency was found to be $f_{lo} - f_{if}$ = 4.53202 GHz. **e.** Extracted Rabi frequencies ($\Omega_R/2\pi$) as functions of $A_{if}$ for the on and off states. When the AQFP mixer is on, $\Omega_R/2\pi$ increases monotonically with $A_{if}$; however, only a residual response is observed when the mixer is off. **f.** Residual ratio of the Rabi frequency as a function of $A_{if}$, highlighting the digital controllability and suppression of residual microwaves.

It should be noted that the dependence of $\Omega_R/2\pi$ on $A_{if}$ when the AQFP mixer is on (Fig. 3e) is markedly non-linear, whereas the reference drive scheme using room-temperature electronics exhibits a linear scaling under the identical qubit conditions. This nonlinearity arises from the nonlinear relationship between the input and output signals of the AQFP mixer[26], rather than from the qubit's characteristics.

## Quantum gate demonstrations

Furthermore, we demonstrate single-qubit gates using the AQFP mixer to evaluate the effect of the AQFP mixer on the coherence characteristics of the transmon qubit. The pulse parameters to achieve $\pi/2$ and $\pi$ pulses (i.e., $X_{\pi/2}$ and $X_\pi$ gates) were initially extracted from the Rabi oscillations in Fig. 3c and then fine-tuned by an error amplification protocol[45], resulting in $f_{lo}$ = 8.0 GHz and $f_{if}$ = 3.470171 GHz, i.e., drive frequency of $f_{lo} - f_{if}$ = 4.529829 GHz. Standard single-qubit characterization[46] was then performed by applying $\pi/2$ and $\pi$ pulses to the transmon qubit with the AQFP mixer, thereby measuring the energy relaxation time ($T_1$), Ramsey dephasing time ($T_2^*$), and echo coherence time ($T_2^e$). For comparison, $T_1$ and $T_2^e$ were also measured using a reference drive scheme based on room-temperature electronics, in which drive pulses were applied to the qubit via the same port used for $V_{read}$ and calibrated in the same way as the AQFP-based scheme. The AQFP-based and reference experiments were performed in the same cooldown to ensure identical experimental conditions.

Figures 4a through 4c show the excited-state population ($P_1$) vs. time for energy relaxation measurements, Ramsey interferometry, and Hahn echo experiments, respectively. The red markers represent measurement results using the AQFP mixer, the blue markers represent those using the reference scheme, and the black lines represent the fitting curves for the red markers. Figures 4a and 4c show that the measurement results using the AQFP mixer agree well with those using the reference scheme: $T_1 = 25.3$ μs and $T_2^e = 17.0$ μs using the AQFP mixer, whereas $T_1 = 26.1$ μs and $T_2^e = 17.6$ μs using the reference scheme. Figure 4b shows that the interference oscillations in the Ramsey interferometry change with tiny detuning magnitudes ($\Delta f$) of 0.05 MHz, 0.34 MHz, and 1.01 MHz, where $\Delta f = (f_{lo} - f_{if}) - f_{qubit}$ ($f_{lo}$ and $f_{qubit}$ are fixed to 8.0 GHz and 4.529829 GHz, respectively). This means the frequency of the microwave pulses produced by the AQFP mixer can be precisely adjusted by varying $f_{if}$ at the 300 K stage.

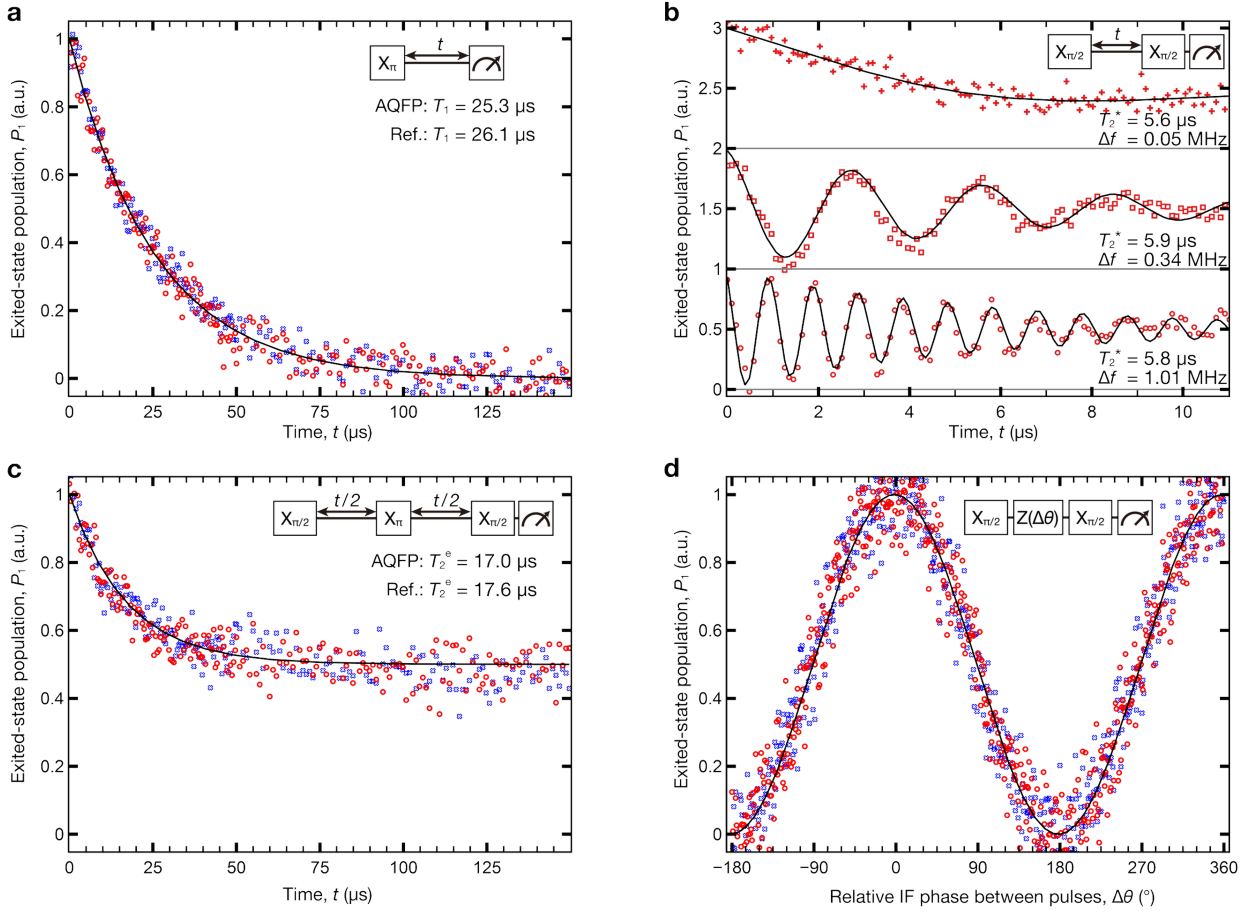

**Fig. 4 | Single-qubit gates using an AQFP mixer and a reference drive scheme.** The red markers represent measurement results using the AQFP mixer, the blue markers represent those using the reference scheme, and the black lines represent the fitting curves for the red markers. **a.** Energy relaxation measurements. $T_1$ was 25.3 μs and 26.1 μs using the AQFP mixer and reference scheme, respectively. **b.** Ramsey interferometry for different detuning magnitudes ($\Delta f$). $T_2^*$ with the AQFP mixer was 5.6 μs, 5.9 μs, and 5.8 μs for $\Delta f$ of 0.05 MHz, 0.34 MHz, and 1.01 MHz, respectively. **c.** Hahn echo experiments. $T_2^e$ was 17.0 μs and 17.6 μs using the AQFP mixer and reference scheme, respectively. **d.** Ramsey interferometry for demonstrating virtual Z gates. The clear interference oscillation indicates arbitrary phase tunability. Taken together, the above measurement results verify coherent single-qubit gates using the AQFP mixer, without observable degradation in coherence or controllability.

In addition to X gates, we executed virtual Z gates by changing the relative IF phase ($\Delta\theta$) between consecutive π/2 pulses in Ramsey interferometry[27]. Figure 4d shows measurement results for $P_1$ vs. $\Delta\theta$, where the clear interference oscillation indicates that the qubit state can be rotated around the equator of a Bloch sphere without applying additional pulses. The above measurement results verify that the AQFP mixer has the capability to execute both X and virtual Z gates, and that the presence of the AQFP mixer does not induce additional decoherence to the transmon qubit at the resolution level of the above experiments. We intend to perform further assessment with gate-fidelity benchmarking in the future.

## Discussions

Here, we compare the performance of various qubit controllers to clarify the impact and benefits of the AQFP QC-VZ. Table 1 summarizes the performance comparison, with the far right column showing the specifications of the AQFP QC-VZ. Item "Power dissipation" denotes the power dissipation of the controllers themselves, excluding the heat load due to wiring, for simplicity. Item "Cable count" represents the number of cables between the 3 K and 10 mK stages, as we assume that these cables are the most difficult to reduce and ultimately limit scalability. Items "X gate" and "Z gate" represent the capabilities to execute X and Z gates, respectively. Item "Memory" represents whether a cryogenic

**Table 1 | Comparison of qubit controllers**

| Technology | Cryo-CMOS | Cryo-CMOS | SFQ | ERSFQ | AQFP |
|---|---|---|---|---|---|
| Circuit | Microwave-pulse generator | Microwave demultiplexer | Pulse-train generator | Pulse demultiplexer | Microwave-pulse generator |
| Power dissipation | 12 mW/qubit | 180 nW/qubit | 1.6 µW/qubit | 2.25 nW/qubit | 111 pW/qubit |
| Operating temperature | 3 K | 10 mK | 3 K | 10 mK | 10 mK |
| Multiplexing | FDM + TDM | TDM | | TDM | µWM + TDM |
| Cable count[a] | $N_{qubit}/32$ | $\sim\log_2(N_{qubit})$ | $N_{qubit}$ | $\sim\log_2(N_{qubit})$ | $\sim N_{qubit}/4000$ |
| Gate parallelism[b] | $N_{qubit}/16$ | 1 | $N_{qubit}$ | 1 | $\sim N_{qubit}$ |
| X gate | ✓ | ✓ | ✓ | ✓ | ✓ |
| Z gate | ✓ | ✓ | ✓ | ✓ | ✓ |
| Memory[c] | ✓ | | | | |
| Reference | 8 | 11 | 14 | 17 | This study |

$N_{qubit}$: number of qubits; FDM: frequency-division multiplexing; TDM: time-division multiplexing; µWM: microwave multiplexing.
[a]Cable count between the 10 mK and 3 K stages, excluding the cables for two-qubit gates and readout, for simplicity.
[b]Maximum number of qubits that can be manipulated in parallel.
[c]Cryogenic memory for storing pulse sequences.

memory for storing pulse sequences is integrated. Overall, Table 1 highlights that superconductor-based qubit controllers exhibit high energy efficiency, while cryo-CMOS-based controllers achieve high functionality, such as memory integration[8–10,12], and that multiplexer-type controllers operate with a small number of cables at the expense of gate parallelism. Most importantly, the AQFP QC-VZ operates with power dissipation orders of magnitude lower than that of the other controllers. Furthermore, the AQFP QC-VZ achieves superior hardware efficiency thanks to the combination of µWM and TDM, i.e., multiple qubits can be manipulated with a small number of cables and high gate parallelism. Currently, the AQFP QC-VZ lacks cryogenic memories for storing pulse sequences, which apply digital input currents (e.g., $I_{in1}$ through $I_{in3}$ in Fig. 1b) to the AQFP mixers to switch on/off each microwave pulse, and the capability to execute two-qubit gates. Therefore, it will be crucial to enhance the circuit functionality of the AQFP QC-VZ while considering which technology is most compatible with AQFPs among various types of cryogenic memories[47–49] and two-qubit gates[50–52].

During the experiments shown in Figs. 3 and 4, the LO power was fixed to 10 dBm at 300 K, which increased the temperature of the 10 mK stage from 10.8 mK to 29.7 mK. This temperature increase should be attributed to Joule heating in the wiring (i.e., 0-dB attenuators, coaxial cables, connectors, and a printed circuit board) between the mixing chamber plate and the AQFP chip, since the power dissipation of the AQFP mixer is negligibly small (~100 pW). According to a supplementary test using a heater, this temperature rise corresponds to a heat load of 13.6 µW, suggesting an attenuation of 2.7 dB in the wiring (see Supplementary Information for details). This heat load is not fundamental to AQFP-based qubit control and can be reduced by lowering the LO power with a low-current-drive AQFP design[36] and resonator-mediated AQFP drive. Also, the attenuation can be reduced by replacing the coaxial cables with superconducting cables.

## Conclusions

We reported the AQFP QC-VZ, an advanced AQFP-based qubit controller. The AQFP QC-VZ can apply both X and virtual Z gates to multiple qubits in parallel, with low power dissipation (111 pW per qubit). Moreover, the number of cables between the 300 K and 10 mK stages can be significantly reduced by combining µWM and TDM. The architectural aspects of the AQFP QC-VZ were validated by demonstrating microwave manipulation capabilities of an AQFP QC-VZ chip at 4.2 K. Furthermore, AQFP-based qubit control was verified by demonstrating single-qubit gates using an AQFP mixer at the 10 mK stage. The single-qubit characterization revealed that the AQFP mixer did not induce observable decoherence to the qubit chip within the present measurement resolution. The above results indicate the feasibility of scalable qubit control using AQFP logic.

## Methods
### AQFP mixer

Extended Data Fig. 1 shows a detailed diagram of the AQFP mixer used in this study, comprising paired AQFPs (A and B), an impedance matching network (IMN), and a BPF. The AQFPs generate microwave pulses ($V_a$ and $V_b$) on the load line by mixing an LO signal ($I_{lo}$) and an IF signal ($I_{if}$) based on the nonlinearity of the AQFPs. The relative phase (0 or π) between $V_a$ and $V_b$ is determined by the digital input currents ($I_{in}$ and $I_{fix}$), and the interference between the pulses enables switching: the microwave output ($V_{out}$) is on when $I_{in}$ and $I_{fix}$ have the same logical value and off otherwise. To derive as large a power as possible from the AQFPs, the impedance of the load line is designed to be a small value of 2 Ω (i.e., $R_s = 2$ Ω). The IMN converts the impedance from 2 Ω to 50 Ω, and the BPF removes unwanted frequency components. Consequently, the AQFP mixer generates and modulates microwave pulses with a frequency of $f_{lo} - f_{if}$ and phase of $-\theta_{if}$. More details regarding the AQFP mixer can be found in the literature[26].

## AQFP design

The AQFPs in the AQFP mixers were designed with 40-µA Josephson junctions. The physical layout designs of the AQFPs were verified by extracting each inductance and coupling coefficient using InductEx[53]. The $\beta_c$ values were 25 and 12 for the AQFPs in the numerical simulations and those in the experiments, respectively.

## Power calculation

In numerical simulations to estimate the output power and power dissipation of the AQFP mixer, a 50 Ω terminating resistor was connected to the output port of the mixer. The output power was calculated as the product of the current through and the voltage across the terminating resistor. The power dissipation was calculated as the product of the currents through the shunt resistors of the Josephson junctions and $R_s$ and the corresponding voltages across them.

## Data availability

The data that support the findings of this study are available from the corresponding author upon reasonable request.

## References

1.  Google Quantum AI and Collaborators *et al.* Quantum error correction below the surface code threshold. *Nature* https://doi.org/10.1038/s41586-024-08449-y (2024) doi:10.1038/s41586-024-08449-y.
2.  Fowler, A. G., Mariantoni, M., Martinis, J. M. & Cleland, A. N. Surface codes: Towards practical large-scale quantum computation. *Phys. Rev. A* **86**, 032324 (2012).
3.  Cho, A. No room for error. *Science* **369**, 130–133 (2020).
4.  Google Quantum AI *et al.* Suppressing quantum errors by scaling a surface code logical qubit. *Nature* **614**, 676–681 (2023).
5.  Krinner, S. *et al.* Engineering cryogenic setups for 100-qubit scale superconducting circuit systems. *EPJ Quantum Technol.* **6**, 2 (2019).
6.  Tian, J. J. *et al.* High-density wiring solution for 500-qubit scale superconducting quantum processors. *Review of Scientific Instruments* **96**, 104709 (2025).
7.  Kawabata, S. Integration and resource estimation of cryoelectronics for superconducting fault-tolerant quantum computers. Preprint at https://doi.org/10.48550/arXiv.2601.03922 (2026).
8.  Van Dijk, J. P. G. *et al.* A scalable cryo-CMOS controller for the wideband frequency-multiplexed control of spin qubits and transmons. *IEEE J. Solid-State Circuits* **55**, 2930–2946 (2020).
9.  Chakraborty, S. *et al.* A cryo-CMOS low-power semi-autonomous transmon qubit state controller in 14-nm FinFET technology. *IEEE J. Solid-State Circuits* **57**, 3258–3273 (2022).
10. Kang, K. *et al.* A 40-nm cryo-CMOS quantum controller IC for superconducting qubit. *IEEE J. Solid-State Circuits* **57**, 3274–3287 (2022).
11. Acharya, R. *et al.* Multiplexed superconducting qubit control at millikelvin temperatures with a low-power cryo-CMOS multiplexer. *Nat Electron* **6**, 900–909 (2023).
12. Underwood, D. *et al.* Using cryogenic CMOS control electronics to enable a two-qubit cross-resonance gate. *PRX Quantum* **5**, 010326 (2024).
13. Leonard, E. *et al.* Digital coherent control of a superconducting qubit. *Phys. Rev. Applied* **11**, 014009 (2019).
14. Howe, L. *et al.* Digital control of a superconducting qubit using a Josephson pulse generator at 3 K. *PRX Quantum* **3**, 010350 (2022).
15. Shen, H., Takeuchi, N., Yamanashi, Y. & Yoshikawa, N. Amplitude-controllable microwave pulse generator using single-flux-quantum pulse pairs for qubit control. *Supercond. Sci. Technol.* **36**, 095010 (2023).
16. Liu, C. H. *et al.* Single flux quantum-based digital control of superconducting qubits in a multichip module. *PRX Quantum* **4**, 030310 (2023).
17. Jordan, C. *et al.* A quantum computer controlled by superconducting digital electronics at millikelvin temperature. *Nature Electronics* https://doi.org/10.1038/s41928-026-01576-6 (2026) doi:10.1038/s41928-026-01576-6.
18. Likharev, K. K. & Semenov, V. K. RSFQ logic/memory family: A new Josephson-junction technology for sub-terahertz-clock-frequency digital systems. *IEEE Trans. Appl. Supercond.* **1**, 3–28 (1991).
19. Mukhanov, O. A. Energy-efficient single flux quantum technology. *IEEE Trans. Appl. Supercond.* **21**, 760–769 (2011).
20. Takeuchi, N., Ozawa, D., Yamanashi, Y. & Yoshikawa, N. An adiabatic quantum flux parametron as an ultra-low-power logic device. *Supercond. Sci. Technol.* **26**, 035010 (2013).
21. Takeuchi, N., Yamae, T., L. Ayala, C., Suzuki, H. & Yoshikawa, N. Adiabatic quantum-flux-parametron: A tutorial review. *IEICE Trans. Electron.* **E105.C**, 251–263 (2022).
22. Takeuchi, N., Yamae, T., Ayala, C. L., Suzuki, H. & Yoshikawa, N. An adiabatic superconductor 8-bit adder with $24k_BT$ energy dissipation per junction. *Applied Physics Letters* **114**, 042602 (2019).
23. Keyes, R. W. & Landauer, R. Minimal energy dissipation in logic. *IBM J. Res. & Dev.* **14**, 152–157 (1970).
24. Likharev, K. K. Classical and quantum limitations on energy consumption in computation. *Int J Theor Phys* **21**, 311–326 (1982).
25. Koller, J. G. & Athas, W. C. Adiabatic switching, low energy computing, and the physics of storing and erasing information. in *Workshop on Physics and Computation* 267–270 (IEEE, Dallas, TX, 1992). doi:10.1109/PHYCMP.1992.615554.
26. Takeuchi, N., Yamae, T., Yamashita, T., Yamamoto, T. & Yoshikawa, N. Microwave-multiplexed qubit controller using adiabatic superconductor logic. *npj Quantum Inf* **10**, 53 (2024).
27. McKay, D. C., Wood, C. J., Sheldon, S., Chow, J. M. & Gambetta, J. M. Efficient Z gates for quantum computing. *Phys. Rev. A* **96**, 022330 (2017).
28. Noroozian, O. *et al.* High-resolution gamma-ray spectroscopy with a microwave-multiplexed transition-edge sensor array. *Appl. Phys. Lett.* **103**, 202602 (2013).
29. Ullom, J. N. & Bennett, D. A. Review of superconducting transition-edge sensors for x-ray and gamma-ray spectroscopy. *Supercond. Sci. Technol.* **28**, 084003 (2015).
30. Delport, J. A., Jackman, K., Roux, P. L. & Fourie, C. J. JoSIM—Superconductor SPICE simulator. *IEEE Trans. Appl.*

*Supercond.* **29**, 1300905 (2019).
31. Krantz, P. *et al.* A quantum engineer's guide to superconducting qubits. *Applied Physics Reviews* **6**, 021318 (2019).
32. Boykin, P. O., Mor, T., Pulver, M., Roychowdhury, V. & Vatan, F. A new universal and fault-tolerant quantum basis. *Information Processing Letters* **75**, 101–107 (2000).
33. Fang, E. S. & Van Duzer, T. A Josephson Integrated Circuit SImulator (JSIM) for SUperconductive Electronics Applications. in (1989).
34. Stewart, W. C. Current-voltage characteristics of Josephson junctions. *Applied Physics Letters* **12**, 277–280 (1968).
35. McCumber, D. E. Effect of ac impedance on dc voltage-current characteristics of superconductor weak-link junctions. *Journal of Applied Physics* **39**, 3113–3118 (1968).
36. Yamae, T. *et al.* Rapid single-flux-quantum and adiabatic quantum-flux-parametron cell libraries using a 1 kA/cm2 niobium fabrication process. *Sci Rep* **15**, 41429 (2025).
37. Irimatsugawa, T. *et al.* Study of Nb and NbN resonators at 0.1 K for low-noise microwave SQUID multiplexers. *IEEE Trans. Appl. Supercond.* **27**, 2500305 (2017).
38. Suzuki H., Takeuchi N. & Yoshikawa N. Development of the wideband cryoprobe for evaluating superconducting integrated circuits. *IEICE Transactions on Electronics (Japanese Edition)* **J104-C**, 193–201 (2021).
39. Kohjiro, S. *et al.* White noise of Nb-based microwave superconducting quantum interference device multiplexers with NbN coplanar resonators for readout of transition edge sensors. *Journal of Applied Physics* **115**, 223902 (2014).
40. Martinis, J. M. *et al.* Decoherence in Josephson qubits from dielectric loss. *Phys. Rev. Lett.* **95**, 210503 (2005).
41. Hornibrook, J. M. *et al.* Cryogenic control architecture for large-scale quantum computing. *Phys. Rev. Applied* **3**, 024010 (2015).
42. Naaman, O., Abutaleb, M. O., Kirby, C. & Rennie, M. On-chip Josephson junction microwave switch. *Applied Physics Letters* **108**, 112601 (2016).
43. Graninger, A. L. *et al.* Microwave switch architecture for superconducting integrated circuits using magnetic field-tunable Josephson junctions. *IEEE Trans. Appl. Supercond.* **33**, 1501605 (2023).
44. Lee, A. C. *et al.* Power dissipation measurement in RQL digital logic. *IEEE Trans. Appl. Supercond.* **33**, 1302809 (2023).
45. Lazăr, S. *et al.* Calibration of drive nonlinearity for arbitrary-angle single-qubit gates using error amplification. *Phys. Rev. Applied* **20**, 024036 (2023).
46. Burnett, J. J. *et al.* Decoherence benchmarking of superconducting qubits. *npj Quantum Inf* **5**, 54 (2019).
47. Nguyen, M.-H. *et al.* Cryogenic memory architecture integrating spin hall effect based magnetic memory and superconductive cryotron devices. *Sci Rep* **10**, 248 (2020).
48. Golod, T., Morlet-Decarnin, L. & Krasnov, V. M. Word and bit line operation of a 1 × 1 μm$^2$ superconducting vortex-based memory. *Nat Commun* **14**, 4926 (2023).
49. Medeiros, O. *et al.* A scalable superconducting nanowire memory array with row–column addressing. *Nat Electron* https://doi.org/10.1038/s41928-025-01512-0 (2026) doi:10.1038/s41928-025-01512-0.
50. Li, R. *et al.* Realization of high-fidelity CZ gate based on a double-transmon coupler. *Phys. Rev. X* **14**, 041050 (2024).
51. Jin, X. Y. *et al.* Superconducting architecture demonstrating fast, tunable high-fidelity *CZ* gates with parametric control of *ZZ* coupling. *Phys. Rev. Applied* **24**, 064026 (2025).
52. Xu, P., Zhang, H. & Wu, S. Protocols for i SWAP gates using a fixed coupler driven by two microwave pulses. *Phys. Rev. Applied* **23**, 034036 (2025).
53. Fourie, C. J. Full-gate verification of superconducting integrated circuit layouts with InductEx. *IEEE Trans. Appl. Supercond.* **25**, 1300209 (2015).

## Acknowledgements

This paper was partially based on results obtained from projects, JPNP16007 and JPNP20004, commissioned by the New Energy and Industrial Technology Development Organization (NEDO), Japan. This study was supported by JST FOREST (Grant No. JPMJFR212B) and JST CREST (Grant No. JPMJCR1676). The circuits were fabricated in the Superconducting Quantum Circuit Fabrication Facility (Qufab) of the AIST.

## Author contributions

HM designed the qubit samples, performed the qubit control experiments, and analyzed the measurement results. AT and WR fabricated the qubit samples. YZ designed the cryogenic packages. TY and JT supported the theoretical aspects. NY supported circuit design environments. NT conceived the idea, designed the AQFP circuits, and performed the numerical simulations and experiments regarding the AQFP circuits. NT and HM wrote the manuscript with contributions from all authors.

## Competing interests

The authors declare no competing interests.

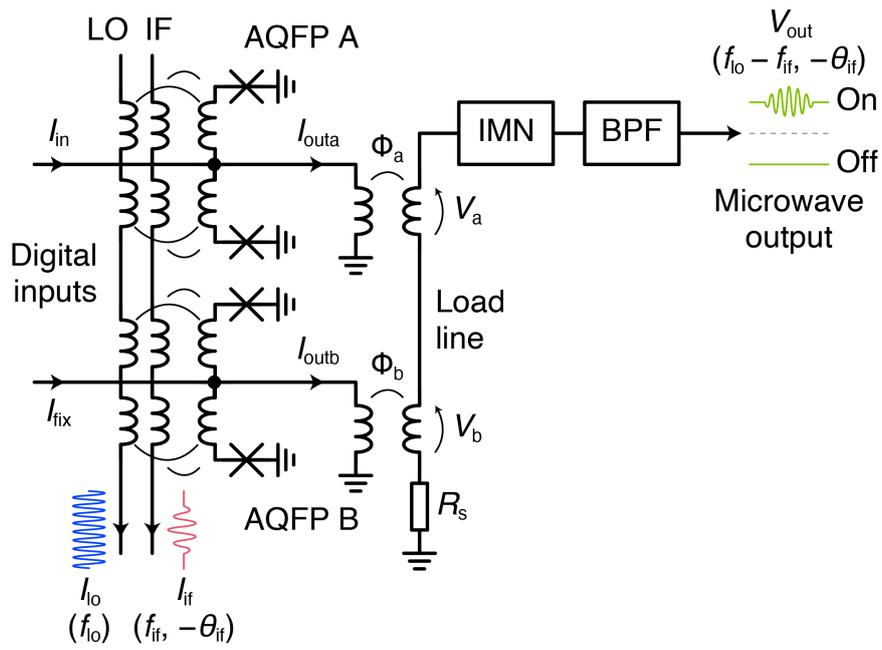

**Extended Data Fig 1 | AQFP mixer.**